\documentclass[12pt]{amsart}
\usepackage{amssymb,amscd,color}
\usepackage{amsmath}
\usepackage{easybmat}
\usepackage{graphics}
\usepackage{graphicx}

\newtheorem{theorem}{Theorem}[section]
\newtheorem{lemma}[theorem]{Lemma}

\newtheorem{corollary}[theorem]{Corollary}
\newtheorem{example}[theorem]{Example}
\newtheorem{simul}[theorem]{Simulation}
\newtheorem{remark}{Remark}
\newtheorem {algo}{Algorithm}

\newcommand{\R}{\hbox{\ensuremath{\mathbb{R}}}}

\newcommand {\set}{\Lambda }

\newcommand{\beq}{\begin {equation}}
\newcommand{\eeq}{\end{equation}}

\numberwithin{equation}{section}

\begin{document}

\title{Sequential adaptive compressed sampling via Huffman codes}

\author[Akram Aldroubi]{Akram~Aldroubi} \address{Department of Mathematics\\
Vanderbilt University\\ 1326 Stevenson Center\\ Nashville, TN 37240}
\email[Akram Aldroubi]{akram.aldroubi@vanderbilt.edu}

\author[Haichao Wang]{Haichao~Wang}

\email[Haichao Wang]{haichao.wang@vanderbilt.edu}

\author[Kourosh Zarringhalam]{Kourosh~Zarringhalam}

\email[Kourosh Zarringhalam]{Kourosh.Zarringhalam@vanderbilt.edu}

 \thanks{ The research of A.~Aldroubi is supported in part by NSF Grant DMS-0807464.}

\maketitle

\begin{abstract}
%\boldmath
In this paper we introduce an information theoretic approach and use techniques from the theory of Huffman codes to construct  a sequence of  binary sampling vectors to determine a sparse signal. Unlike the standard approaches, ours is adaptive in the sense that each sampling vector depends on the previous sample results. We prove that the expected total cost (number of measurements and reconstruction combined) we need for an $s$-sparse vector in $\R^n$ is no more than  $s\log n + 2s$.
\end{abstract}


% \fi

\section{Introduction}
Let $x$ be a vector in $\R^n$ and assume that the vector $x$ has at most $s<< n$ nonzero components, denoted $\|x\|_0\le s$.  In compressed sampling, the goal is to determine a set of linear functionals   ({\em the sampling functions}) and associated reconstruction algorithms such that, if the set of functionals is applied to $x$, then the reconstruction algorithms will allow us to find $x$ from the values of the functionals (measurements of $x$)  in a computationally tractable way which is stable (i.e., that  should produce a good approximation even when $x$ is not $s$-sparse) and robust  to noise (i.e. that should produce a good approximation to $x$ when the measurements (sampling) are corrupted by additive noise).  There is a trade off between the number of sampling vectors we need to acquire $x$ and the computational cost of the reconstruction algorithm which determines $x$ from the samples.  

Some of the earlier work uses an $\ell_1$ minimization for finding $x$ from a set of samples $\{y_i=\langle a_i, x\rangle, \; i=1,\dots,m\}$ where the $a_i$s are vectors in $\R^n$, see e.g., \cite {CR06,CRT06,Dev07,Don06,GN03,Tro04}. Letting $y=(y_1,\dots,y_m)^t$ and $A$ be the $m\times n$ matrix whose rows are the vectors $a_i$, the $\ell_1$ minimization approach finds the unknown $s$-sparse vector $x$ (i.e., $\|x\|_0\le s$) by  solving the constrained minimization problem
\begin {equation}
\label {ell1min}
\min \|z\|_{\ell_1}, \quad Az=y.
\end{equation}
The solution to the minimization problem produces $x$ as its unique solution provided $A$ is an appropriate sampling matrix, e.g., $A$ satisfying the \emph {Restricted Isometric Property} (RIP).  The current methods for constructing matrices satisfying the needed RIP are probabilistic and produce matrices with $m=O(s\log(n/2s))$ rows (i.e., we need $m=O(s\log(n/2s))$ sampling vectors). The deterministic construction of a matrix $A$ satisfying the needed RIP property for arbitrary $n (m<<n)$ is still open. Finding $x$ via the $\ell_1$ minimization problem involves linear programing with $n$ variables and $m$ constraints which can be computationally expensive. 

Matching pursuits (OMP, ROMP, CoSaMP. etc., see \cite {NT08,NV09} and the reference therein) form another class of sampling and reconstruction algorithms for finding an $s$-sparse signal $x$ from the measurements $y=Ax$.  These algorithms are iterative in nature and view the column of $A$ as a dictionary for the reconstruction of $x$. At every step of the reconstruction algorithms, a  certain number of columns (column indices) are chosen in order to minimize an equation of the form $\|r_n-Ax_n\|_2$, where $r_n=y-Ax_{n-1}$ is a residual from the previous step. The most performant of these type of  reconstruction algorithms is relatively fast, stable, and robust to noise. However, similar to the $\ell_1$ minimization algorithms, it requires a sampling matrix $A$ satisfying the needed RIP. As before only probabilistic methods  are known, and they produce matrices with $m=O(s\log(n/2s))$ rows that satisfy the RIP with high probability. 

Other approaches for compressed sampling are combinatorial.  A sampling matrix $A$ is constructed using bipartite graphs, such as  expander graphs, and the reconstruction finds  an unknown $s$-sparse vector $x$ using binary search methods, see e.g. \cite {BGIKS08, GKMS03, DWB05, SBB06b, SBB06a,  GSTV06, GSTV07,  XH07} and the references therein. Typically, the matrix $A$ has binary entries. There exist fast algorithms for finding the solution $x$ from the measurements (typically sublinear). However, the construction of $A$ is still difficult to produce. 

There are emerging new approaches for adaptive methods in compressed sampling. One approach uses a Bayesian method combined  a gaussian  model for the measurements and a Laplacian  model for the sparsity \cite {JXC08}. The sampling vectors are chosen by minimizing a differential entropy.   Another approach uses is a type of binary search algorithm that uses back projection and block adaptive sampling to focus on the possible nonzero component (see \cite{HCN09} the references therein).

The new approach we present is information theoretic and uses tools from the theory of Huffman codes to develop a deterministic  construction of a sequence of binary sampling vectors $a_\set$, i.e., the entries of $a_\set$ consist of $0$ or $1$.  Moreover, unlike standard approaches, the sampling procedure is adaptive.  We assume that the signal $x\in \R^n$ is an instance of a vector random variable $X=(X_1,\dots,X_n)^t$ and we construct the $i$-th row $a_i$ of $A$ using the sample $y_{i-1}=\langle a_{i-1}, x\rangle$. Our goal is to make the average number of samples needed to determine a signal $x$ as small as possible, and to make the reconstruction of $x$  from those samples as fast as possible. 
We take advantage of the probability distribution of the random vector $X$ to minimize the average number of samples needed to uniquely determine the unknown signal $x$.  In our method, rather than constructing a fixed set of sampling vectors forming the rows of a single sampling matrix $A$ for all possible signals, we construct $s$ sequences of sampling vectors. Each sequence focuses on finding exactly one nonzero component of the signal $x$. 

It is remarkable that the expected total cost of the combined sampling and reconstruction algorithms of an $s$-sparse vector $x$ is no more than $s\log n+2s$. If no information is available about the probability distribution of the random vector $X$, we can always assume that we have a uniform distribution, in which case the total cost of the combined sampling and reconstruction algorithms of an $s$-sparse vector $x$ is equal to $s\log n+2s$ even if the uniform assumption is erroneous.

This paper is organized as follows: In Section \ref  {NP} we introduce the basic notation and definitions for sparse random vectors and trees. The new notion of Huffman tree and Huffman sampling vectors
are introduced in Section \ref {th1} and \ref {th2}. In Sections \ref {th3} and \ref {th4}, we describe an information theoretic method for finding $s$-sparse vectors in $\R^n$. In Section \ref {th5} we describe a variation  for finding $s$-sparse vectors from noisy measurements. Section \ref {ES} is devoted to examples, simulations, and testing of the algorithms on synthetic data. 
\section {notation and preliminaries}
\label {NP}
In this section we introduce the necessary notations and preliminaries needed in subsequent sections. 
\subsection {Sparse random vectors}
\label{SRV}
\begin {enumerate}
\item We will use the notation $X = (X_1, \dots, X_n)^t $ to denote  a vector of $n$ random variables. An instance $x \in \R^n$ of $X$ will be called a  \emph {signal}.   

\item We will say that a signal $x \in \R^n$ has sparsity $s\le n$ if $x$ has at most $k$ nonzero components ($\|x\|_0\le s$). 

\item Let $\Omega=\{1, \dots, n\}$ be the set of all indices, and  $\set \subset \Omega$ be a subset of indices. Then $P_{\set}=P(\set)$ will denote the probability of  $X$ having nonzero components exactly at coordinates indexed by $\set$, i.e.,  $P_\set=Pr\{X_i \neq 0, i \in \set \; ;\; X_i = 0, i  \in \set^c \} $. Here $\set^c$ denotes the complement of $\set$. 

\item The pair $(X,P)$ will be used to denote the random vector $X$ together with the probability mass function $P$ on the sample space $2^\Omega=2^{\{1,\dots,n\}}$. Thus obviously $\sum_\set P_\set=1$.

\item We will say that a random vector $X$ is $s$ sparse if $P_\set=0$ for all $\set$ with cardinality strictly larger than $s$, i.e., $ \#(\set)>s$ implies $P_\set=0$.

\item We will need the probabilities $q_\set=q(\set)=Pr(E_\set)$ for the events $E_\set=\{ X_i\ne 0,  \text { for some }  i \in \set \} $, which is the probability that at least one of the components of $X$ with index in $\set$ is nonzero. Note that $q$ can be computed from $P$ by
\begin{equation}
\label{q}
q_\set=\sum\limits_{\eta \cap \set\neq \emptyset}P_\eta.
\end{equation}
\end {enumerate} 
 \subsection {Trees}
 \label {Tree}
 \begin {enumerate}
\item  We consider finite \emph {full binary trees}  in which every node has zero or two children.
 
\item The \emph{root} of the tree is the node with no parent nodes. 

\item A \emph {leaf} is a node with no children. 
 
\item A \emph {left (right) subtree} of a node $v$ in a rooted binary tree is the tree whose root is the left (right) child of this node $v$.  

\item The set of all nodes that can be reached from the root by a  path of length $L$ are said to be at \emph {level} $L$. 
\end {enumerate}
 \subsection {Other notations} 
 \label {ON}
 \begin {enumerate}
\item The notation $\chi_{\set}$ will denote the characteristic function of a set $\set$, i.e., $\chi_\set(i)=1$ for $i \in \set$ and $\chi_\set(i)=0$ for $i \notin \set$.
\item For a set $\set$, $|\set|$ will denote its cardinality. 
 \end{enumerate}
 
 \section{Theory}
 \label {Th}
 In this section we describe our approach explicitly.

 \subsection {Huffman tree}
 \label {th1}
Let $(X,P)$ be a $s$-sparse random vector $X$ in $\R^n$ together with the probability mass function $P$ on the sample space $2^\Omega=2^{\{1,\dots,n\}}$ the set of all subsets of $\Omega=\{1,\dots,n\}$.  We define a Huffman tree to be a binary tree whose leaves are the  sets ${\{1\}},\dots,{\{n\}}$. We associate probabilities $q_{\{1\}},\dots,q_{\{n\}}$ to these nodes respectively. The Huffman tree is constructed from the leaves to the root as follows: Suppose that the nodes at the $i$-th step are ${\set_1},\dots,{\set_s}$. Let $i$ and $j$ be such that

\begin{tabular}{lll}
 $q_{\set_i}$ & $=$ & $\underset{1 \leq \lambda \leq s}{\min} 
q_{\set_\lambda},$\\
 $q_{\set_j}$ & $=$ & $\underset{1 \leq \lambda \leq s, \lambda \neq i
}{\min} q_{\set_\lambda}.$
\end{tabular}$\;$\\\\
Then, the nodes at the $(i+1)$-th step are obtained by replacing 
${\set_i}$ and ${\set_j}$ with ${\set_i \cup \set_j}$ in 
the list of the nodes at the $i$-th step and the probability associated 
to the node ${\set_i \cup \set_j}$, will be $q_{\set_i \cup 
\set_j}$, i.e., at each step the two nodes with smallest probabilities are combined. An illustrative example is shown below.
\begin{example}
\label {ex1}
Assume that $X \in \R^4$ is a $2$-sparse random vector with probability 
mass function $P$ defined by: $P_\emptyset = 0.02, P_{\{1\}} = 0.07,  
P_{\{2\}} = 0.05,  P_{\{3\}} = 0.03,  P_{\{4\}} = 0.1,  P_{\{1,2\}} = 
0.31,  P_{\{1,3\}} = 0.2,  P_{\{1,4\}} = 0.03,  P_{\{2,3\}} = 0.06,  
P_{\{2,4\}} = 0.12,  P_{\{3,4\}} = 0.01$. The nodes at the first step 
are: ${\{1\}},\dots,{\{4\}}$. Using equation \ref{q}, we get that:

$$
q_{\{1\}} =  P_{\{1\}} + P_{\{1,2\}} + P_{\{1,3\}} + P_{\{1,4\}} = 0.61.
$$$\;$\\
Similarly, $q_{\{2\}} = 0.54, q_{\{3\}} = 0.3,$ and $q_{\{4\}} = 0.26$. 
Therefore the nodes at the second step are ${\{1\}}, {\{2\}}, 
{\{3,4\}}$. Also $q_{\{3,4\}} = 0.55$ and hence the nodes at the third 
step are ${\{1\}}, {\{2,3,4\}}$. The root note is ${\{1, 2, 
3,4\}}$ with probability $q_{\{1, 2, 3,4\}} = 1$. This completes the 
Huffman tree (See Figure 1).

\begin{figure}
\begin{center}
\label {exmpletree1}
    \includegraphics[width=2.5in, height=2.0in]{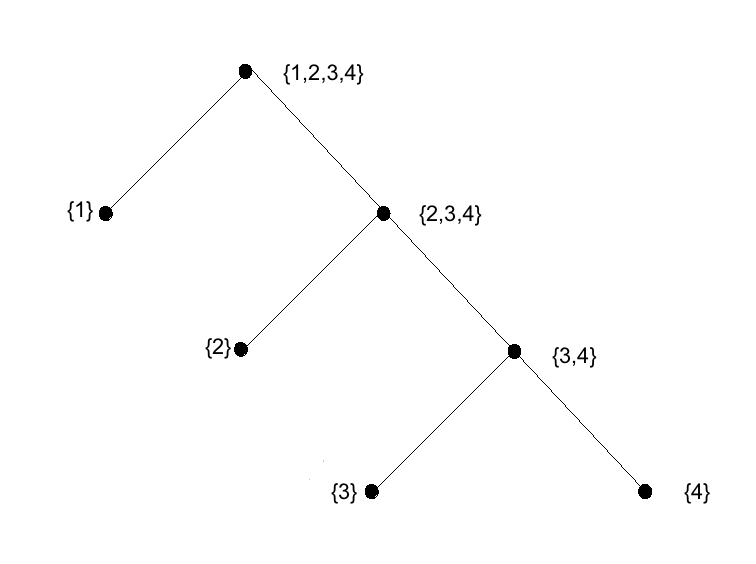}
\end{center}
\caption{Huffman Tree for the $2$-sparse random vector $X \in \R^4$.}
\end{figure}
\end{example}

\subsection{Huffman sampling vectors}
\label {th2}
In this section we introduce Huffman sampling vectors. Let  $(X,P)$ be as in the previous section. Our goal is to find a signal $x$ which is an instance of $X$ using, on average, a minimum number of samples. The  average  number of samples is measured by 
\begin {equation}
\label {AL}
L(X,P)=\sum_\set P_\set \ell_\set,  
\end{equation}
where $\ell_\set$ is the number of samples for finding one nonzero component of $x$ whose components $x_i\ne 0$ for $i\in \set$ and $x_i= 0$ for $i\in \set^c$.

As mentioned before, we first focus on finding one nonzero component, if it exists. Other nonzero components are then iteratively found one at a time in a similar manner until all of the nonzero components are exhausted.  

We first construct a Huffman tree associated with the random vector $X$. Each node $\set$ which is not a leaf has two children ${\set_1}$ and ${\set_2}$.  Note that $ \set_1\cap \set_2=\emptyset$ and $\set_1 \cup \set_2 =\set$.  We denote $\ell_{{\set_1}}=q_{\set_1}(\log|\set_1|+1)+(1-q_{\set_1})(\log|\set_2|+1)$ and $\ell_{\set_2}=q_{\set_2}(\log|\set_2|+1)+(1-q_{\set_2})(\log|\set_1|+1)$.
 We associate a sampling vector to such nodes $\set$ by:

\begin {equation}
\label {sv}
a_{\set}=
\left\{
\begin{tabular}{lcr}
$\chi_{\set_1} $ & & \text{if}\ $l_{\set_1} \leq l_{\set_2},$\\
$\chi_{\set_2} $ & & \text{if}\  $l_{\set_1} > l_{\set_2},$
\end{tabular} \right.
\end {equation}
 i.e., for  $l_{\set_1} \leq l_{\set_2}$,  we have $a_\set(i)=1$ for $i \in \set_1$ and $a_\set(i)=0$ for $i \in \Omega-\set_1$, and for  $l_{\set_1} > l_{\set_2}$ we have $a_\set(i)=1$ for $i \in \set_2$ and $a_\set(i)=0$ for $i \in \Omega-\set_2$.
 
The choice of the sampling vector in \eqref {sv} can be seen as follows:  Since our goal is to find the nonzero component as quickly as possible,   it seems that we would need $a_\Lambda=\chi_{\set_i} $ for the set $\set_i$ with the highest probability $q_{\set_i}$, $i=1,2$. However, the set $\set_i$ with the highest probability $q_{\set_i}$ may also have a large number of elements. Thus the choice should be a compromise between the size of the set and its probability. This particular choice of $a_{\set}$ will be apparent from the theorems and their proofs below.

\subsection{ Determination of a sparse vector $x$ using Huffman sampling vectors.}
\label {th3}
Let $x$ be an $s$-sparse signal in $\R^n$ which is an instance of $(X,P)$ (we will write $x\sim (X,P)$). We  make the additional assumption that  the conditional probability 
$Pr(\sum_{i\in\set}X_i\ne 0| X_i\ne 0, i\in \set)=0$ holds for any $\set\subset \Omega$ (recall that $\Omega=\{1.\,\dots,n\}$).  This is a natural condition if the random variables $X_i$, $i=1,\dots,n$, in the random vector $X$ do not have a positive  mass concentration except possibly at zero.  
\subsubsection {Finding a nonzero component}
\label {FNC1}
Algorithm \ref {alg1} below is used  to find  the position and the corresponding value of one of the nonzero  components of $x$  (if any). 
\begin {algo} $\text { }$
\label {alg1}
\begin {enumerate}

\item Initialization: $\set=\Omega$;

\item Repeat until $|\set|=1$
%\begin {enumerate}

 $  {\quad               } $  if $\langle a_\set, x \rangle \ne 0$, $\set=\set_1$
 	
$  {\quad               } $  else $\set=\set_2$

\noindent end repeat
\item Output the (only) element $t_1 \in \set$ 
\item Output $x_{t_1}=\langle \chi_{\{t_1\}},x\rangle$ 
\end {enumerate}
\end {algo}
\begin {remark}   $\text{}$
\begin {enumerate}
\item [(i)] If the vector $x=0$, then the algorithm will find an output $x_{t_1}=0$, otherwise  it will output the value $x_{t_1}$ of one of the nonzero components of $x$ and its index  $t_1$. 
\item  [(ii)] Note that the last inner product output in (4) is not always necessary since we have all the information needed to find $x_{t_1}$ from the samples and the value of $t_1$. However, this would involve solving a linear system of equations obtained from the sampling scheme. Thus this one  extra sample can be considered as reconstruction step.
\item  [(iii)] Note that the sampling vectors depend on the instance $x$, i.e., the sampling vectors are adaptive. 
\item  [(iv)] The number of possible sampling vectors is equal to the number of nodes in the Huffman tree, but only a subset of these vectors is used to determine a nonzero component of a given instance vector $x$. 
\item [(v)] If $P(X=0)>0$, then we choose the first sampling  vector $a= (1,1,\dots,1)$. If $\langle a, x \rangle = 0$ we are done. Otherwise we proceed with Algorithm \ref {alg1}.  
\end {enumerate}
\end {remark}
The first observation is that Algorithm 1 is optimal for $1$-sparse vectors. This should not be a surprise since the algorithm was inspired by the theory of Huffman codes. We have
\begin {theorem} \label{1sparse}
Given a 1-sparse vector $x \sim (X,P)$ in $\R^n$. Then the average number of samples $L_1(X,P)$ needed to find $x$ using Algorithm \ref {alg1} is less than or equal to the average number of samples $L_A(X,P)$ for finding $x$ using any algorithm $A$ with binary sampling vectors.
\end {theorem}
\begin{proof}
Let $E=\{e_i\}^n_{i=1}$ be the canonical basis for $\R^n$. We first note that for any sampling algorithm with binary sampling vectors, the number of samples required to determine any nonzero multiple $\alpha e_i$ of $e_i$ ($\alpha \in \mathbb{R}$) is equal to the number required to determine $e_i$. Hence for the remainder of this proof we will assume that $x$ is binary, i.e., $x$ is one of the canonical vectors $e_i$. 
Since any sampling vector $a$ from an algorithm $A$ is binary, the inner product $\langle a, x\rangle$ is either zero or one, i.e., binary.  For each $e_i$, $i \in \Omega=\{1,\dots,n\}$, a binary algorithm uses a sequence of vectors $\{a^i_1,\dots,a^i_{\ell_i}\}$ where $\ell_i$ is the number of sampling vectors needed to determine $e_i$. We associate to each $e_i$ the binary sequence  $c^i=y^i_1y^i_2\dots y^i_{\ell_i}$ where $y^i_j= \langle a^i_j, e_i\rangle$. In this way each canonical vector $e_i$ is associated to a unique sequence $c^i$, $c^i\ne c^j$ if $i \ne j$.  Hence $c^i$ is a binary code for the vectors of the canonical basis of $\R^n$.  The code is a prefix (also known as instantaneous code, i.e., no code is a prefix of any other code \cite {CT91}) since a binary algorithm terminates after finding the nonzero component (which correspond to the shorter code). Since $x$ is $1$ sparse, we have that $q_{i}=P_{i}$ which is the probability of   component $i$ being nonzero and all other components being zero. Hence for each algorithm $A$ with binary sampling vectors, we associate a prefix code for $E=\{e_i\}^n_{i=1}$. Consequently, the average  number  of samplings required in this algorithm $A$ is the same as the average length of the code:  $L_A(X,P)=\sum\limits_{i=1}^n\ell_iP_{i}.$ From the construction, Algorithm \ref {alg1} is associated with  the Huffman code whose average length is the shortest. Hence the average number of samples $L_1(X,P)\le L_A(X,P)$ for any $A$.
 \end{proof}

The $1$-sparse case is a special case. It is optimal because the sampling scheme can be associated exactly with the Huffman codes and we have that $q_\set=P_\set$ and $\sum_iq_{\{i\}}=\sum_iP_{\{i\}}=1$.  In fact, the choice of $a_\set$ in \eqref {sv} can be chosen to be either $\chi_{\set_1}$ or $\chi_{\set_2}$ independently of the values of $\ell_{\set_1},\ell_{\set_2}$.  However, for the general $s$-sparse case, we do not have $q_\set=P_\set$ anymore, and the sampling vectors cannot be associated with Huffman codes directly. Thus for the $s$-sparse case, the average number of sampling vectors is not necessarily optimal and we need to estimate this number to have confidence in the algorithm. 

From the construction of the Huffman tree in Section \ref {th1}, it is not difficult to see that there is at most one node $\set$ (called \emph{special node}) with children $\set_1$ and $\set_2$ such that $(\frac{1}{2}-q_{\set_1})(\frac{1}{2}-q_{\set_2})< 0$, i.e., except for possibly the special node, all other nodes have the property that $q_{\set_1}, q_{\set_2}$ are either both  larger than $1/2$ or both smaller than $1/2$.  Thus,  a Huffman tree can at most have one special node. We have the following lemmas:

\begin{lemma}\label{nspecial}
For any fixed node $\set$ with children $\set_1$ and $\set_2$, if $\set$ is not a special node, then $\min\{\ell_{\set_1}, \ell_{\set_2}\}\leq \log|\set|$.
\end{lemma}
\begin{proof}
Without loss of generality, we assume that $|\set_1|\leq|\set_2|$. Hence $\frac{|\set_1|}{|\set|}\leq\frac{1}{2}$ and $\frac{|\set_2|}{|\set|}\geq\frac{1}{2}$.

Consider the function $f_q(x)=x^q(1-x)^{1-q}$ for $x \in [0,1]$ and $q\in [0,1]$. Easy computations show that
\begin {equation}
\label {Case1}
f_{q}(x)\leq\frac{1}{2},\  \text{for} \ q\leq\frac{1}{2}\  \text{and} \ x\geq\frac{1}{2},
\end{equation}
and 
\begin{equation}
\label {Case2}
f_{q}(x)\leq\frac{1}{2},\  \text{for} \ q\geq\frac{1}{2}\  \text{and} \ x\leq\frac{1}{2}.
\end{equation}
Since $\set$ is not a special node, we have that $(q_{\set_1}-\frac{1}{2})(q_{\set_2}-\frac{1}{2})\geq0$.

If $q_{\set_1}\leq\frac{1}{2}$ and $q_{\set_2}\leq\frac{1}{2}$, then using the fact that $\frac{|\set_2|}{|\set|}\geq\frac{1}{2}$ and \eqref {Case1}, we get $f_{q_{\set_2}}(\frac{|\set_2|}{|\set|})\leq\frac{1}{2},$ that is
$$(\frac{|\set_2|}{|\set|})^{q_{\set_2}}(1-\frac{|\set_2|}{|\set|})^{1-q_{\set_2}}\leq\frac{1}{2},$$
which implies that 
$$2|\set_2|^{q_{\set_2}}|\set_1|^{1-q_{\set_2}}\leq|\set|,$$
after taking the log function on both sides, we have 
$$\ell_{\set_2}\leq\log|\set|.$$

 A similar calculation for the case $q_{\set_1}\geq\frac{1}{2}$ and $q_{\set_2}\geq\frac{1}{2}$ yields
$$\ell_{\set_1}\leq\log|\set|.$$

In either case, we have that $\min\{\ell_{\set_1}, \ell_{\set_2}\}\leq \log|\set|$.
\end{proof}

For general node, we have the following,
\begin{lemma}\label{special}
For any fixed node $\set$ with children $\set_1$ and $\set_2$, we have that $\min\{\ell_{\set_1}, \ell_{\set_2}\}\leq \log|\set|+1$.
\end{lemma}
\begin {proof}
For any $x\in [0,1]$ and any $q\in [0,1]$ we have that $f_q(x)=x^q(1-x)^{1-q}\leq 1$. We use this inequality for $x=\frac{|\set_1|}{|\set|}$ and $q=q_{\set_1}$ or $x=\frac{|\set_2|}{|\set|}$ and $q=q_{\set_2}$ to obtain the result.
\end {proof}

\begin{lemma}\label{nlemma}
Given a nonzero $s$-sparse vector $x \sim (X,P)$ in $\R^n$. If the Huffman tree associated with $x$ has no special node, then the average number of samples $L$ needed to find the position of one nonzero component of $x$ using Algorithm \ref {alg1} is at most $\log n$.
\end{lemma}
\begin{proof}
We will use induction on $n$ to prove this lemma. Suppose $\set=\{1, \dots, n\}$ and $\set$ has children $\set_1$ and $\set_2$.

For $n=2$, $|\set|=2$, we only need one vector $\chi_{\set_1}$ or $\chi_{\set_2}$ to determine the position of one nonzero component of $x$. Hence the lemma holds trivially for this case.

Now assume the lemma is true for $n=k, k-1, \dots, 2$, we want to show it is also true for $n=k+1$.

If $|\set|=k+1$, we must have $|\set_1|\leq k$ and $|\set_2|\leq k$. Without loss of generality, suppose $\ell_{\set_1}\leq\ell_{\set_2}$. By Algorithm \ref {alg1}, $a_{\set}=\chi_{\set_1}$. Then with probability $q_{\set_1}$, we have $\langle a_\set, x \rangle \ne 0$, in which case we need (on average) another $L_{\set_1}$ sampling vectors. With probability $1-q_{\set_1}$, we have $\langle a_\set, x \rangle=0$, and we need (on average) another $L_{\set_2}$ sampling vectors. By the induction hypothesis, we have $L_{\set_1}\leq \log|\set_1|$ and $L_{\set_2}\leq \log|\set_2|$.

Since by assumption the tree has no special node,using Lemma \ref{nspecial} we deduce that the average number of sampling vectors we need is 
\begin{align*}
L &=q_{\set_1}(1+L_{\set_1})+(1-q_{\set_1})(1+L_{\set_2}) \\
&\leq q_{\set_1}(1+\log|\set_1|)+(1-q_{\set_1})(1+\log|\set_2|)\\
&= \ell_{\set_1}\leq\log|\set|=\log k.
\end{align*}
\end{proof}

We are now ready to find an upper bound on the average number of sampling vectors needed for finding the position of one nonzero component in an $s$-sparse signal using Algorithm \ref {alg1}. Denoting by $T_{\set}$ the subtree with $\set$ as the root,  we have 
\begin{theorem}\label{main}
Given a nonzero $s$-sparse vector $x \sim (X,P)$ in $\R^n$ the average number of samples $L$ needed to find the position of one nonzero component of $x$ using Algorithm \ref {alg1} is at most $\log n+1$.
\end{theorem}
\begin{proof}
We will use induction.  The lemma holds trivially for $n=2$.

Now assume the lemma is true for $n=k, k-1, \dots, 2$, we want to show it is also true for $n=k+1$.

If $|\set|=k+1$, we must have $|\set_1|\leq k$ and $|\set_2|\leq k$. Without loss of generality, suppose $\ell_{\set_1}\leq\ell_{\set_2}$. By Algorithm \ref {alg1}, the average number $L$ of sampling vectors needed is  
\begin {equation}
\label {ANSV}
L =q_{\set_1}(1+L_{\set_1})+(1-q_{\set_1})(1+L_{\set_2}).
\end{equation}
Since  the Huffman tree can have at most one special node, we consider three cases:

Case(1): If the root of the tree $\set= \{1, \dots, n\}$ is a special node and  $\set_1$, $\set_2$ are its children. Then the subtrees $T_{\set_1}$ and $T_{\set_2}$ have no special nodes. Thus by Lemma \ref{nlemma}, we have that $L_{\set_1}\leq\log|\set_1|$ and  $L_{\set_2}\leq\log|\set_2|$.

From Lemma \ref{special}, $\ell_{\set_1}\leq\log|\set|+1$. Thus we have that
\begin{align*}
L &=q_{\set_1}(1+L_{\set_1})+(1-q_{\set_1})(1+L_{\set_2}) \\
&\leq q_{\set_1}(1+\log|\set_1|)+(1-q_{\set_1})(1+\log|\set_2|)\\
&= \ell_{\set_1}\leq\log|\set|+1=\log (k+1)+1.
\end{align*}

Case(2): If $\set= \{1, \dots, n\}$ is not a special node and the subtree $T_{\set_1}$ has no special node, from Lemma \ref{nlemma}, we have $L_{\set_1}\leq\log|\set_1|$. Since $|\set_2|\leq k$, from the induction hypothesis, we have $L_{\set_2}\leq \log|\set_2|+1$.

From Lemma \ref{nspecial}, $\ell_{\set_1}\leq\log|\set|$. Thus we have that
\begin{align*}
L &=q_{\set_1}(1+L_{\set_1})+(1-q_{\set_1})(1+L_{\set_2}) \\
&\leq q_{\set_1}(1+\log|\set_1|)+(1-q_{\set_1})(1+\log|\set_2|+1)\\
&= \ell_{\set_1}+1-q_{\set_1}\leq\log|\set|+1=\log (k+1) +1.
\end{align*}

Case(3): If $\set= \{1, \dots, n\}$ is not a special node and the subtree $T_{\set_2}$ has no special node, then the same computation as in Case (2)  gives $L\leq\log (k+1)+1$.
\end{proof}

\subsubsection {Iterative step for finding another nonzero component}
\label {IS}
 For every subset $\omega\subset \Omega$ with $|\omega|<s$ we let $P^\omega_\set=P^\omega(\set)$ denote the conditional probabilities
 $$Pr\big\{X_i \neq 0, i \in \set \; \text{and}\; X_i= 0, i  \in \Omega-\{\set\cup\omega\} | X_{i}\neq 0, \; i \in \omega \big\},$$ for any $\set \subset \Omega-\omega$.

Similar to Section \ref {th1}, we let $q^\omega_\set=q^\omega(\set)=Pr(E^\omega_\set)$ for the events $E^\omega_\set=\{ X_i\ne 0,  \text { for some }  i \in \set | X_i\neq 0, i \in \omega\} $, which is the conditional probability that at least one of the components of $X$ with index in $\set\subset \Omega-\omega$ is nonzero given that $X_i\neq 0$ for $i \in \omega$. Using the same procedure as in Section \ref {th1}, build a Huffman tree  with leaves $\{i\}, \; i \in \Omega-\omega$ with probabilities $q_\set^\omega$. Note that this tree has $n-|\omega|$ leaves.
 
As above (see \eqref {sv}), we assign a sampling vector $a^\omega_{\set}\in \R^n$ to every node ${\set}$ which is not a leaf. Note that from the construction of $a^\omega_\set$ we have that $a^\omega_\set(i)=0$ for $i \in \omega$.

Let $k< s$ be the number of nonzero components of $x$ that are found and let $\omega=\{t_1,\dots,t_k\}$ be the set of corresponding indices. 
The algorithm for finding the $(k+1)$-th nonzero component of $x$ (if any) is essentially the same as in Algorithm \ref {alg1}. 
\subsection{Algorithm for finding all nonzero components of $x$}
\label {th4}

The general algorithm for finding the $s$-sparse vector $x \sim (X,P)$ which is an instance of the $s$-sparse random vector $(X,P)$, can now be described as follows:

\begin {algo} \text {}
\label {GenAlg}

 \noindent Initialization: s=1; $\omega=\emptyset$;\\
 Repeat until $k>s$ or $x_{t_k}=0$
 \begin {enumerate}
\item $\set=\Omega-\omega$;
\item repeat until $|\set|=1$\\
$  {\quad               } $  if $\langle a^{\Omega-\omega}_\set, x \rangle \ne 0$, $\set=\set_1$;\\
 $  {\quad               } $  	else $\set=\set_2$;\\
 end repeat
 \item Output the (only) element $t_k \in \set$;
 \item Output $x_{t_k}=\langle \chi_{\{t_k\}},x\rangle$;
\item $\omega=\omega\cup \{t_k\}$;
\item $k=k+1$;
\end {enumerate}
end repeat
\end{algo}
Algorithm \ref{GenAlg} repeats Algorithm \ref {alg1} at most $s$ times and adds one extra sample to determine the value of each nonzero component once its position is known. Thus, as a corollary of Theorem \ref{main} we immediately get

\begin {corollary} 
\label {GenAS}
Given a nonzero $s$-sparse vector $x \sim (X,P)$ in $\R^n$, the average number of sampling vectors $L$ needed to find all nonzero components of $x$ using Algorithm \ref {GenAlg} is at most $s\log n+2s$.
\end {corollary}

\begin{remark} \text{}

\begin {enumerate}
\item [(i)] Corollary \ref {GenAS} states that the upper bound on the expected total cost (number of measurements and reconstruction combined) that we need for an $s$-sparse vector in $\R^n$ using Algorithm \ref{GenAlg} is no more than  $s\log n + 2s$.
\item [(ii)] If the probability distribution $P$ is uniform then the combined cost  of the measurements and reconstruction is exactly $s\log n + 2s$.
\end{enumerate}
\end {remark}

%\subsection {Computational complexity and optimality}
\subsection{Noisy measurements}
\label {th5}
In practice, the measurements $\{y_i\}$ maybe corrupted by noise. Typically the noise is modeled as  additive and uncorrelated : $y_\set=\langle x, a_\set\rangle + \eta_\set$ (see \cite {CW08}). For this case the condition $Az=y$ in \eqref {ell1min} for the $\ell_1$ minimization technique is modified to $\|Az-y\|_2\le \epsilon$ where $\epsilon$ is of the same order as the standard deviation  $\sigma_\eta$ of the noise. With this modification, the $\ell_1$ minimization technique yields  a minimizer $x^\star$ satisfying  $\|x^\star-x\|_2\le C\epsilon$ where $C$ is a constant independent of $x$. Similar modifications are made for the other techniques, e.g., $\ell_q$ minimization (see \cite {FL09}). 

Similarly, our algorithm needs to be modified accordingly to deal with noisy measurements case. Algorithm \ref {GenAlg} can be modified by changing the statement $\langle a^{\Omega-\omega}_\set, x \rangle \ne 0$ to the statement $|\langle a^{\Omega-\omega}_\set, x \rangle| > T$, where the threshold $T$ is of the same order as the standard deviation of $\eta$.  

Consider the model $Y=X+\eta$ where the signal $X\sim N(0, \sigma_X)$ and the noise $\eta\sim N(0, \sigma_\eta)$. Then $Y\sim N(0, \sqrt{\sigma_X^2+\sigma_\eta^2})$. We set the threshold in Algorithm \ref {GenAlg} to be $T=E(|\eta|)=\frac{2\sigma_\eta}{\sqrt{2\pi}}$, and consider a measure of error (for one sample)  given by  the probability $$p(e)=P\big(|Y|< T\ \text{and}\ |X|\geq T|\big)+P\big(|Y|\geq T\ \text{and}\ |X|< T|\big)$$

After easy computation, we have that 
$$p(e)=\text{erf}(\frac{\sigma_\eta}{\sqrt{\pi}\sigma_X})+\text{erf}(\frac{\sigma_\eta}{\sqrt{\pi}\sigma_Y})-\text{erf}(\frac{\sigma_\eta}{\sqrt{\pi}\sigma_X})\text{erf}(\frac{\sigma_\eta}{\sqrt{\pi}\sigma_Y}),$$
where $\text{erf}(x)=\frac{2}{\sqrt{\pi}}\int_0^xe^{-t^2}dt$ is the error function.

Using the Taylor series, we obtain
$$p(e)=\frac{4t}{\pi}-\frac{4t^2}{\pi^2}+o(t^2),$$
where $t=\frac{\sigma_\eta}{\sigma_X}$ is the ratio of the standard deviations of the noise and the signal. Thus, for a relatively large signal to noise ratio,  $t$ will be small and  we get that the probability of at least one error in the sampling-reconstruction for an $s$ sparse vector is bounded above by the quantity  
$$p_s(e)=1-(1-p(e))^{s(\log n+1)}=s(\log n+1)p(e)+o(p(e)^2)\approx s(\log n+1)\frac{4t}{\pi}.$$

It can be seen that $p_s(e)$ is  essentially linear in the sparsity $s$, linear in $t$ and logarithmic in the dimension $n,$ as can also be seen in the simulations below.

%For example if the  $1$-sparse case,  if we assume that the  signal $X$ has a value given by a uniform probability density function in  $[-A,A]$,  then  a threshold of $\epsilon=A/40$ will produce a $5\%$ chance of error. Thus, for this choice of $\epsilon$ the standard deviation of the noise should not exceed $\epsilon$. 

\subsection {Stability and compressible signals}
One of the advantage of the standard compressed sensing methods is that they produce almost optimal results for signal that are not $s$-sparse. For example, if we let $\beta^1_s (x)$ denote the smallest possible error (in the $\ell^1$ norm) that can be achieved by approximating a signal $x \in \R^n$ by an $s$-sparse vector $z$:
\[ 
\beta_s (x):=\inf\{\|x-z\|_1,\|z\|_0\le s\},
\]  
then the vector $x^\star$ solution to the $\ell_1$ reconstruction method \eqref {ell1min} is quasi-optimal in the sense that $\|x-x^\star\|_1\le C\beta_s (x)$ for some constant $C$ independent of $x$. Since for a given $x\in \R^n$, the quantity $\beta^1_s(x)$ is the $\ell_1$ norm of the smallest $n-s$ components of $x$,  the previous result means that if $x$ is not $s$-sparse, then $x^\star$ is close to the $s$-sparse vector $x_s$ whose components are the $s$-largest components of $x$. In particular, if $x$ is sparse, then $x^\star=x$.  

Clearly, our current approach cannot produce similar results since, in its current form, our method does not have any incentive for finding  the largest non-zero components. However $x$ can be decomposed into $x=x_s+(x-x_s)$, and a measurement $y=\langle a_\Lambda,x_s\rangle+ \eta_\Lambda$ where $\eta_\Lambda=\langle a_\Lambda, ,(x-x_s)\rangle$ can be viewed as noise. Thus, a possible modification of the method is to replace $\langle a^{\Omega-\omega}_\set, x \rangle \ne 0$ by $|\langle a^{\Omega-\omega}_\set, x \rangle| > \Delta_k(\set)$ in Algorithm \ref {GenAlg} where $\Delta_k(\set)$ depends on the characteristics of the random variables $X_i$, and $\set$. However, such modification will not be studied in this paper.

\section {Examples and Simulations}
\label {ES}
In this section, we provide some examples and test our algorithm on synthetic data. In the first experiment we use an exponential distribution to generate the position of the nonzero components of the $s$-sparse vectors, and uniform distribution for their values. The signal $x$ is generated by first generating an integer index $i\in [1, n]$ using the exponential pdf  with a mean of $10$, and then constructing the component $x(i)=A(rand-0.5)$, where $rand$ is a random variable with uniform distribution in $[0,1]$. In all the other experiments, we use a uniform probability distribution for both signal and noise. The signal $x$ is generated by first generating an integer index $i\in [1, n]$ with a uniform distribution, and then constructing the component $x(i)=A(rand-0.5)$, where $rand$ is a random variable with uniform distribution in $[0,1]$. This process is repeated $s$ times for $s$ sparse signals. Each  additive noise $\eta_\set$, is generated by $\eta_\set=N*(rand-0.5)$ and added to the measurement $y_\set$. All the experiments are done  using Matlab
7.4 on a Macintosh MacBook Pro  2.16 GHz Intel Core Duo processor 1GB 667 MHz RAM. 
\subsection {Noiseless cases}
\begin {simul}
Our first experiment is a sparse vector $x\sim (X,P)$ in a space of dimension $n=2^{15}$ with an exponential pdf with mean $10$  for the location of the nonzero components and a uniform distribution for the values of the components as described above. We have tested our algorithm with $s=1,3,5,7,9,11,13$. The mean and variance of the number of sampling vectors  needed for the various sparsity $s$ (for the combined sampling and reconstruction) is shown is Table \ref{T2}.
\begin{table}[htdp]
\begin{center}
\begin{tabular}{|c|c|c|c|c|c|c|c|} \hline
s & 1 & 3&5&7&9&11&13\\ \hline
$s\log n$ &15&45&75&105&135&165&195\\ \hline
Mean & 9.11&27.5&46.17&61.47&81.26&97.2&111.88\\  \hline
Var &10.15&16.6&25.5&25.88&31.88&30.94&37.37\\

\hline
\end{tabular}
\end{center}\vspace{2mm}
\caption{Mean and variance of number of sampling vectors as functions of sparsity $s$ for $n=2^{15}=32768.$}
\label{T2}
\end{table}
\end{simul}

\begin {simul}
Our second example is a sparse random vector $X$ in a space of dimension $n=1024$ with a uniform probability distribution. We have tested our algorithm on an example with  $n=1024$, $s=1, 25, 50, 75, 100, 125, 150$. The time for finding the vector $x$ for the various sparsity $s$ is shown in Table \ref {T1}.  It is clear from the table that the methods is  very performant. 

\begin{table}[htdp]
\begin{center}
\begin{tabular}{|c|c|c|c|c|c|c|c|} \hline
s & 1 & 25&50&75&100&125&150\\ \hline
CPU time & 0.0045&0.028&0.049&0.073&0.098&0.144&0.146\\
\hline
\end{tabular}
\end{center}\vspace{2mm}
\caption{CPU time as a function of sparsity $s$ for $n=1024.$}
\label{T1}
\end{table}%gin {example} [nonumiform case]
\end{simul}

\subsection{ Noisy measurements}

\begin {simul} 
In this test we fix the following values: $n=512$, $A=20$, $N=0.1$. For each value of $s$, we construct $100$ $s$-sparse signals in $\R^n$. We test the effect of $s$ on the $\ell^2$ relative error $\frac {\|\hat x- x\|_2} {\| x\|_2}$ (in percent) as a function of the sparsity $s$. The results are displayed in Figure 2. The experiments shows that the  relative error increases linearly with $s$.

\begin{figure}
\begin{center}
\label {Errk1}
    \includegraphics[width=2.5in]{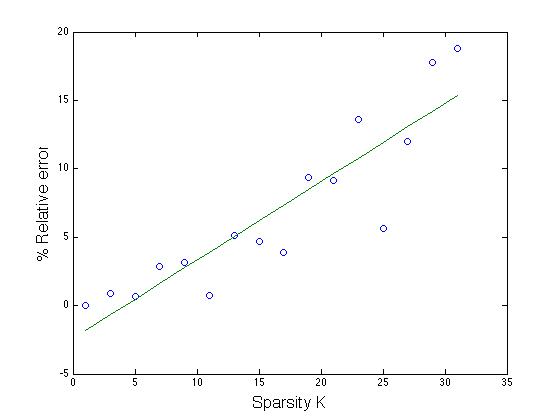}
\end{center}
\caption{Relative $\ell^2$ error of reconstruction from noisy measurements.  }
\end{figure}
\end{simul}

\begin {simul} 
In this test we fix the following values: $n=512$, $A=20$,  $s=16$. We test the effect of $noise$ on the $\ell^2$ relative error $\frac {\|\hat x- x\|_2} {\| x\|_2}$ (in percent) as a function of the value $N$ of the noise. The results are displayed in Figure 3. The experiments suggest that the the relative error increases linearly with $N$.

\begin{figure}
\begin{center}
\label {noiseerr}
    \includegraphics[width=2.5in]{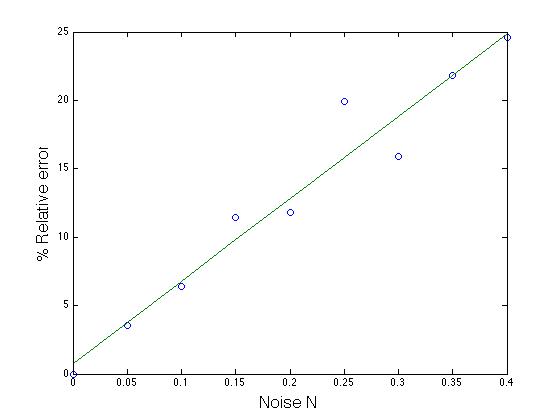}
\end{center}
\caption{Relative $\ell^2$ error of reconstruction from noisy measurements.  }
\end{figure}
\end{simul}

\begin {simul} 
In this test we fix the following values: $A=20$,  $N=0.1$, $s=8$. We test the effect of $n=2^r$ on the $\ell^2$ relative error $\frac {\|\hat x- x\|_2} {\| x\|_2}$ (in percent) as a function of the value $r$. The results are displayed in Figure 4. 
\begin{figure}
\begin{center}
\label {dimerr}
    \includegraphics[width=2.5in]{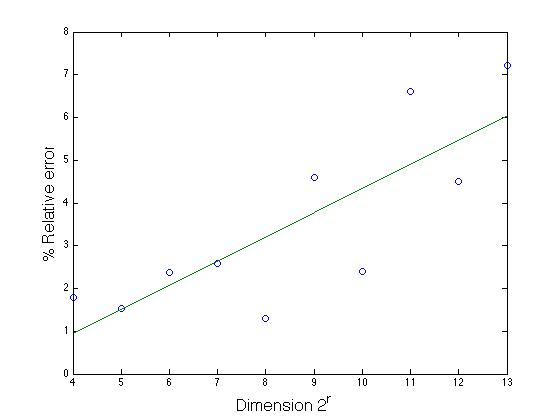}
\end{center}
\caption{Relative $\ell^2$ error of reconstruction from noisy measurements.  }
\end{figure}
\end{simul}

\section {Conclusion}
We have presented an information theoretic approach to compressed sampling of sparse signals. Using ideas similar to those in Huffman coding, we constructed an adaptive sampling scheme for sparse signals. In our scheme, the sampling vectors are binary and their construction  are deterministic and can be produced explicitly for each $n$. Without noise the reconstruction is exact, and the average cost for sampling and reconstruction combined for an $s$ sparse vector is  bounded by $slog(n)+2s$.  We have also shown that the method is also stable in noisy measurements. However, the current method and algorithms are not adapted to  the compressive signals and developments for these cases will be investigated in future research.  We hope that the approach will stimulate further developments and interactions between the area of information theory and compressed sampling.


\begin{thebibliography}{1}
\bibitem[AEB06a]{AEB06b}
M.~Aharon, M.~Elad, and A.M. Bruckstein, \emph{The k-svd: An algorithm for
  designing of overcomplete dictionaries for sparse representation}, IEEE
  Trans. On Signal Processing \textbf{54} (2006), no.~11, 4311 -- 4322.

\bibitem[AEB06b]{AEB06a}
Michal Aharon, Michael Elad, and Alfred~M. Bruckstein, \emph{On the uniqueness
  of overcomplete dictionaries, and a practical way to retrieve them}, Linear
  Algebra Appl. \textbf{416} (2006), no.~1, 48--67.

%\bibitem[AG01]{AG01}
%A.~Aldroubi and K-H. Gr{\"o}chenig, \emph{Non-uniform sampling in
%  shift-invariant space}, Siam Review \textbf{43} (2001), no.~4, 585--620.

\bibitem[BDDW07]{BDDW07}
R.~Baraniuk, M.~Davenport, Ronald~A. DeVore, and M.~Wakin, \emph{A simple proof
 of the restricted isometry property for random matrices}, Preprint, 2007.

\bibitem [BGIKS08] {BGIKS08}
R. Berinde, A. C. Gilbert, P. Indyk, H. Karloff, and M. J. Strauss, 
\emph{Combining geometry and combinatorics: A unified approach to sparse 
signal recovery}, (Preprint, 2008).

%\bibitem[Bow00]{Bow00}
%Marcin Bownik, \emph{The structure of shift-invariant subspaces of
%  ${L}^2(\mathbb{R}^n)$}, Journal of Functional Analysis \textbf{177} (2000),
%  282--309.

%\bibitem[Cas00]{Cas00}
%Peter~G. Casazza, \emph{The art of frame theory}, Taiwanese J. Math. \textbf{4}
%  (2000), no.~2, 129--201.

%\bibitem[Chr03]{Chr03}
%Ole Christensen, \emph{An introduction to frames and {R}iesz basis}, Applied
%  and Numerical Harmonic Analysis, Birkh\"auser, 2003.

\bibitem[CR06]{CR06}
E.~Cand{\`e}s and J.~Romberg, \emph{Quantitative robust uncertainty principles
  and optimally sparse decompositions}, Foundations of Comput. Math. \textbf{6}
  (2006), 227--254.

\bibitem[CRT06]{CRT06}
E.~Cand{\`e}s, J.~Romberg, and Terence Tao, \emph{Robust uncertainty
  principles: Exact signal reconstruction from highly incomplete frequency
  information}, IEEE Trans. on Information Theory \textbf{52} (2006), 489--509.
  
  \bibitem[CW08]{CW08}
E.~Cand{\`e}s and  M.~Wakin, \emph{People hearing without listening: An introduction to compressive sampling}, preprint, 2008.

\bibitem [CT91]{CT91}
T. M. Cover and J. A. Thomas, Elements of Information Theory, Wiley Interscience Publishing, New York, 1991.
 
\bibitem[CT06]{CT06}
E.~Cand{\`e}s and Terence Tao, \emph{Near optimal signal recovery from random
  projections: Universal encoding strategies}, IEEE Trans. on Information
  Theory \textbf{52} (2006), 5406--5425.

\bibitem[DeV07]{Dev07}
Ronald~A. DeVore, \emph{Deterministic constructions of compressed sensing
  matrices}, Preprint, 2007.

\bibitem[Don06]{Don06}
David~L. Donoho, \emph{Compressed sensing}, IEEE Trans. on Information Theory
  \textbf{52} (2006), 1289--1306.

\bibitem[DVB07]{DVB07}
Pier~Luigi Dragotti, M.~Vetterli, and T.~Blu, \emph{Sampling moments and
  reconstructing signals of finite rate of innovation: Shannon meets
  strang-fix}, IEEE Transactions on Signal Processing \textbf{55} (2007),
  1741--1757.
  
 \bibitem  [DWB05] {DWB05} M. F. Duarte, M. B. Wakin, and R. G. Baraniuk. \emph{Fast reconstruction of piecewise smooth signals from random projections}. In Proc. SPARS05, Rennes, France, Nov. 2005.

%\bibitem[EY36]{EY36}
%C.~Eckart and G.~Young, \emph{The approximation of one matrix by another of
%  lower rank}, Psychometrica \textbf{1} (1936), 211 -- 218.
\bibitem [FL09]{FL09}
S. Foucart, M.-J. Lai. \emph{Sparsest solutions of underdetermined linear systems via $\ell_q$-minimization for $0 < q \le 1$}. 
Applied and Computational Harmonic Analysis, 26/3, 395--407, 2009. 

\bibitem [GKMS03] {GKMS03} A. C. Gilbert, Y. Kotidis, S. Muthukrishnan, and M. Strauss, \emph{ One-Pass Wavelet Decompositions of Data
Streams} IEEE Trans. Knowl. Data Eng., \textbf {15} (3) (2003), 541--554.

\bibitem [GSTV06] {GSTV06}A. C. Gilbert, M. J. Strauss, J. A. Tropp, and R. Vershynin. Algorithmic linear dimension reduction in
the $\ell_1$ norm for sparse vectors. Submitted for publication, 2006.
\bibitem [GSTV07] {GSTV07} A. C. Gilbert, M. J. Strauss, J. A. Tropp, and R. Vershynin. One sketch for all: fast algorithms for
compressed sensing. In ACM STOC 2007, pages 237Ð246, 2007.

\bibitem[GN03]{GN03}
R.~Gribonval and M.~Nielsen, \emph{Sparse decompositions in unions of bases},
  IEEE Trans. Inf. Theory \textbf{49} (2003), 3320--3325.

\bibitem[HCN09]{HCN09}
J. Haupt, R. Castro, and R. Nowak, \emph{Adaptive sensing for sparse signal recovery}, Proc. IEEE Digital Signal Processing Workshop and Workshop on Signal Processing Education, Marco Island, FL, 2009, January 2009. 
%\bibitem[Gr{\"o}01]{Gro01}
%K.~Gr{\"o}chenig, \emph{Foundations of time-frequency analysis}, Applied and
%  Numerical Harmonic Analysis, Birkh{\"a}user, 2001.
\bibitem[JXC08]{JXC08}
S. H. Ji and Y. Xue and L. Carin, \emph{Bayesian compressive sensing}, IEEE Transactions On Signal Processing, vol. 56 no. 6 (2008), ppt. 2346 -- 2356
%\bibitem[HJ85]{HJ85}
%R.~Horn and C.~Johnson, \emph{Matrix analysis}, Cambridge University Press,
%  Cambridge, 1985.

%\bibitem[HW96]{HW96}
%Eugenio Hern{\'a}ndez and Guido Weiss, \emph{A first course on wavelets}, CRC
%  Press, Boca Raton, FL, 1996.

\bibitem[LD07]{LD07}
Y.~Lu and M.~N. Do, \emph{A theory for sampling signals from a union of
  subspaces}, IEEE Transactions on Signal Processing, (2007).

\bibitem[MV05]{MV05}
I.~Maravic and M.~Vetterli, \emph{Sampling and reconstruction of signals with
  finite rate of innovation in the presence of noise}, IEEE Transactions on
  Signal Processing \textbf{53} (2005), 2788--2805.
  
  \bibitem [NT08]{NT08} 
D. Needell and J. A. Tropp, \emph {CoSaMP: Iterative signal recovery from incomplete and inaccurate samples},
Applied and Computational Harmonic Analysis, vol. 26, no. 3(2008), pp. 301-321.
  
  \bibitem  [NV09]{NV09} 
 D. Needell and R. Vershynin, \emph{"Uniform Uncertainty Principle and signal recovery via Regularized Orthogonal Matching Pursuit}, Foundations of Computational Mathematics, vol. 9, no. 3(2009), pp. 317-334.

\bibitem[RSV06]{RSV06}
Holger Rauhut, K.~Schass, and P.~Vandergheynst, \emph{Compressed sensing and
  redundant dictionaries}, Preprint, 2006.

\bibitem [SBB06a] {SBB06a} S. Sarvotham, D. Baron, and R. G. Baraniuk. Compressed sensing reconstruction via belief propagation.
Technical Report ECE-0601, Electrical and Computer Engineering Department, Rice University, 2006.
\bibitem [SBB06b] {SBB06b} S. Sarvotham, D. Baron, and R. G. Baraniuk. Sudocodes - fast measurement and reconstruction of sparse
signals. IEEE International Symposium on Information Theory, 2006.
%\bibitem[Sch07]{Sch07}
%E.~Schmidt, \emph{Zur theorie der linearen und nichtlinearen
%  integralgleichungen. i teil. entwicklung willk{\"u}rlichen funktionen nach
%  system vorgeschriebener}, Math. Ann. \textbf{63} (1907), 433--476.

\bibitem[Tro04]{Tro04}
J.~A. Tropp, \emph{Greed is good: Algorithmic results for sparse
  approximation}, IEEE Trans. Inf. Theory \textbf{50} (2004), 2231--2242.

%\bibitem[VMS05]{VMS05}
%R.~Vidal, Y.~Ma, and S.~Sastry, \emph{Generalized principal component analysis
%  (gpca)}, IEEE Transactions on Pattern Analysis and Machine Intelligence
%  \textbf{27} (2005), 1--15.

\bibitem [XH07]{XH07}
W. Xu and B. Hassibi, \emph{ Efficient compressive sensing with determinstic 
guarantees using expander graphs} IEEE Information Theory Workshop, 2007.
\end{thebibliography}
\end{document}